\documentclass[12pt,preprint]{aastex}

\begin{document}

\newcommand{\apol } {$P_{800}$\ }
\newcommand{\apolf} {$P_{800f}$\ }
\title{The Discovery of Diffuse Radio Polarization Structures in the NVSS}
\author{Lawrence Rudnick and Shea Brown}
\affil{Department of Astronomy, University of Minnesota, Minneapolis, MN  55455}
\vskip .5in

\begin{abstract}
We have developed a method for recovering polarization structures from the NRAO Very Large Array Sky Survey (NVSS)  on larger angular scales than the nominal 15 arc minute survey limit.  The technique depends on the existence of smaller scale fluctuations in polarization angle, to which the interferometer is sensitive, while the undetected total intensity of the structures can be arbitrarily large.  We recover the large scale structure of the polarized Milky Way, as seen in single dish surveys, as well as a wide variety of smaller scale galactic and extragalactic features.  We present a brief discussion of the uncertainties and limitations of the reprocessed NVSS polarization survey, a comparison of single-dish and NVSS results, and a sampling of the new polarization structures.  We show a companion feature 1.8~Mpc outside of Abell cluster 3744, apparent  Mpc-scale extensions to the tailed radio galaxy 3C31, a possible new giant galactic loop,and a new bright polarized patch in supernova remnant CTA1.  We note that there is little quantitative information from these detections, and followup investigations would be necessary to measure reliable polarized fluxes and position angles.   Some of the new features discovered in this NVSS reanalysis could provide a foreground for CMB polarization studies, but the internal foreground modeling for the next generation of experiments should have no difficulty accounting for them. 

\end{abstract}

\section{Introduction}
Polarized synchrotron radiation is a powerful diagnostic tool for astrophysical relativistic plasmas.  It provides information on the field direction, degree of order, and in some special situations, the actual field strength.  Polarization's diagnostic power (and its sometimes ambiguous interpretation) for optically thin sources arise from the combination of two factors -- its vector nature, allowing for constructive/destructive interference, and the effects of radiative transfer, primarily Faraday rotation, along the line of sight. 
  
  There is a long history to the use of radio polarizations for structural studies. In our Galaxy, polarized emission allowed the mapping of the large scale structure of the magnetic field including very large angle structures such as Loop 1 \citep{berk}.  More recently, higher resolution images have uncovered structures on a wide range of scales, including the influence of Faraday screens \citep{CGPS, wier,gaens01}.  In addition, discrete objects such as supernova remnants and supershells are highly polarized, allowing study of their magnetic field structures and dynamics \citep{kundu, milne,jones,west}.  Many extragalactic radio sources were found to be significantly polarized, from the pc-scale relativistic jets of blazars and other AGN \citep{wardle71,rud85} to the diffuse lobes of radio galaxies on scales of 100s of kpc \citep{stull}.    In addition to its propitious qualities, polarized synchrotron emission is also important as a foreground for CMB polarization studies \citep{tegmark}.

 Polarized objects like the Mpc-scale peripheral relics or gischt \citep[e.g.,][]{hanisch,kempner} around clusters of galaxies, presumably due to accretion shocks \citep{ens98} were the initial motivation for the current study.   The current work was motivated by a desire to use polarization as a probe to push below the limits of confusion \citep{rud04} to search for signatures of infall into and along the large scale filaments that are part of cosmic structure formation \citep{miniati}. 
 
Although the Stokes parameters Q, U, and I for linearly polarized emission contain a great deal of information, the way in which they are processed and communicated necessarily destroys some of that information. In particular, whether or not a vector background subtraction (in Q and U) or spatial smoothing is done before calculation of the polarized intensity $P=[Q^2+U^2]^{0.5}$ can cause features to appear or disappear in an image. Thus, it is often possible to re-process Stokes images and derive further information than emerged from their initial analysis.  It was this type of thinking that led us to reconsider the polarization information present in the NRAO VLA Sky Survey \citep[NVSS,][]{nvss}.  We note that the proper way to recover diffuse  polarization information unconfused by smaller sources is to combine single dish and interferometer data.  However, since there are angular scales and rotation measure ranges now reachable by the ``allsky'' NVSS        that are unlikely to be duplicated in the near future, it is worth examining what information can be derived from this survey.

The NVSS is a 1.4 GHz survey (combining data from 1364.9 and 1435.1 MHz) conducted from 1993-1997 in the D and DnC configurations (lowest angular resolution) of the VLA, covering the 82\% of the sky which is visible from its latitude. It is reproduced as 2326 4$^o~\times$~4$^o$ images in Stokes parameters I, Q, and U at a resolution of 45".  The rms brightness is $\sim$0.45 mJy/beam (0.14 K), in I, with typical rms values of $\sim$0.29 mJy/beam (0.09\ K) in Q and U at high galactic latitudes.  Faraday rotation between the two NVSS bands eliminates any polarized emission in the images with $\vert$RM$\vert \geq$ 340 rad/m$^2$.  

The information on large scale structures in the NVSS is limited by two effects.  First, the snapshot observations and smallest projected spacings of $\sim$37\ m produce grating lobes at $\sim$18', effectively eliminating the sensitivity to sources larger than that scale.  In addition, the short spacings were further weighted down in the mapping procedure, to remove the large scale  ``pedestal" in the synthesized beam.  Although no detailed study has been done, the NVSS is generally recognized to have little sensitivity to structures $>$15'.   For most extragalactic studies, this provides a great benefit both in total intensity and polarization because it removes the effects of the strong large-scale galactic emission.  However, there are many interesting galactic and extragalactic structures on intermediate scales that do not appear in the original survey.  

It is possible to recover some of this intermediate scale structure from the NVSS linear polarization images.   If a source has structure in Q and U on angular scales to which the interferometer is sensitive, then it will be detected even if the underlying total intensity (I) is much smoother and undetectable by the interferometer.  Q and U often have smaller-scale fluctuations than total intensity because they are sensitive to field disorder, magnetic field direction, and Faraday rotation, none of which affects the total intensity.  The recovery of such polarized features is the purpose of our reanalysis; the NVSS ``all sky" images in total intensity and polarization at 800" resolution are shown in Figure \ref{allsky}, dramatically illustrating the larger scale information present in the latter.  The processing scheme for these images is discussed in detail below.

\subsection{Polarization Image Preview}

\begin{figure}[]
\begin{center}
\includegraphics[width=15cm]{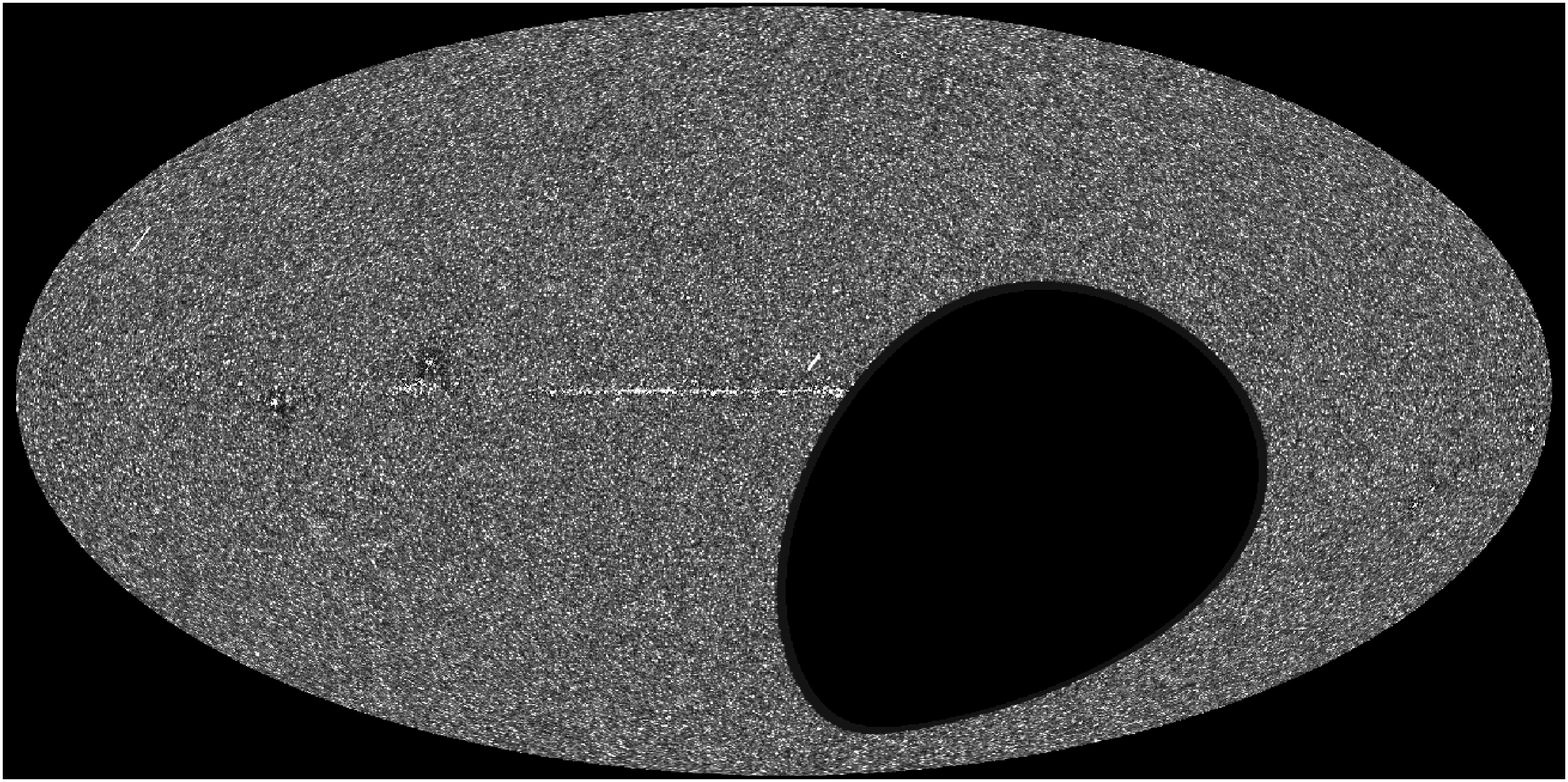}
\vskip 1in
\includegraphics[width=15cm]{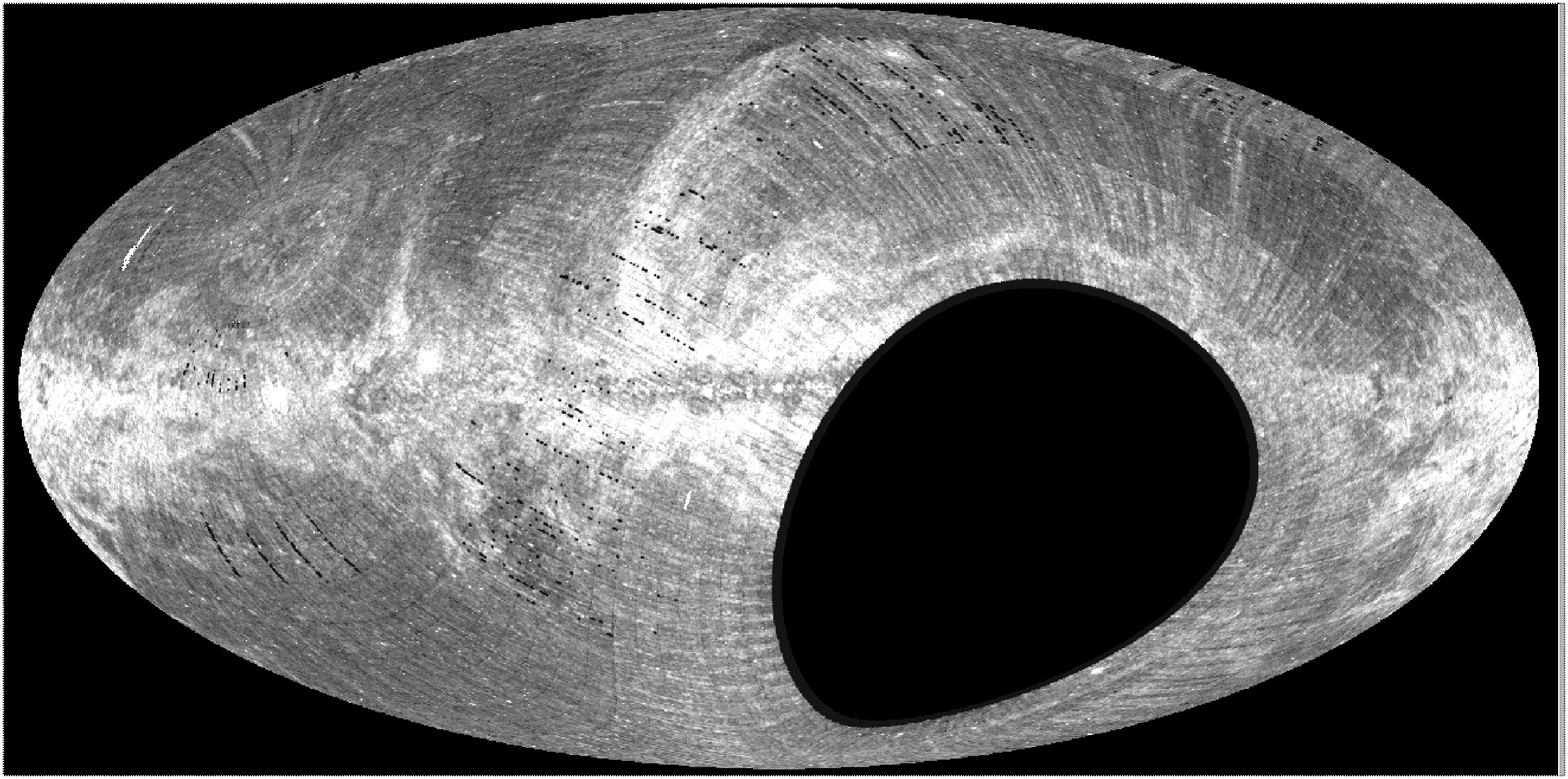}
\end{center}
\caption{All sky AITOFF projections in galactic coordinates of the NVSS, at a resolution of 800". The top image shows the total intensity, with polarized intensity on the bottom.  The obvious striping in the polarized intensity, mostly along lines of declination, are due to variations in the residual instrumental polarization during the survey.  Occasional small black regions is where the residual instrumental polarization was especially high, and the polarization data were flagged for that individual pointing.}
\label{allsky}
\end{figure}

\begin{figure}
\begin{center}
\includegraphics[width=17cm]{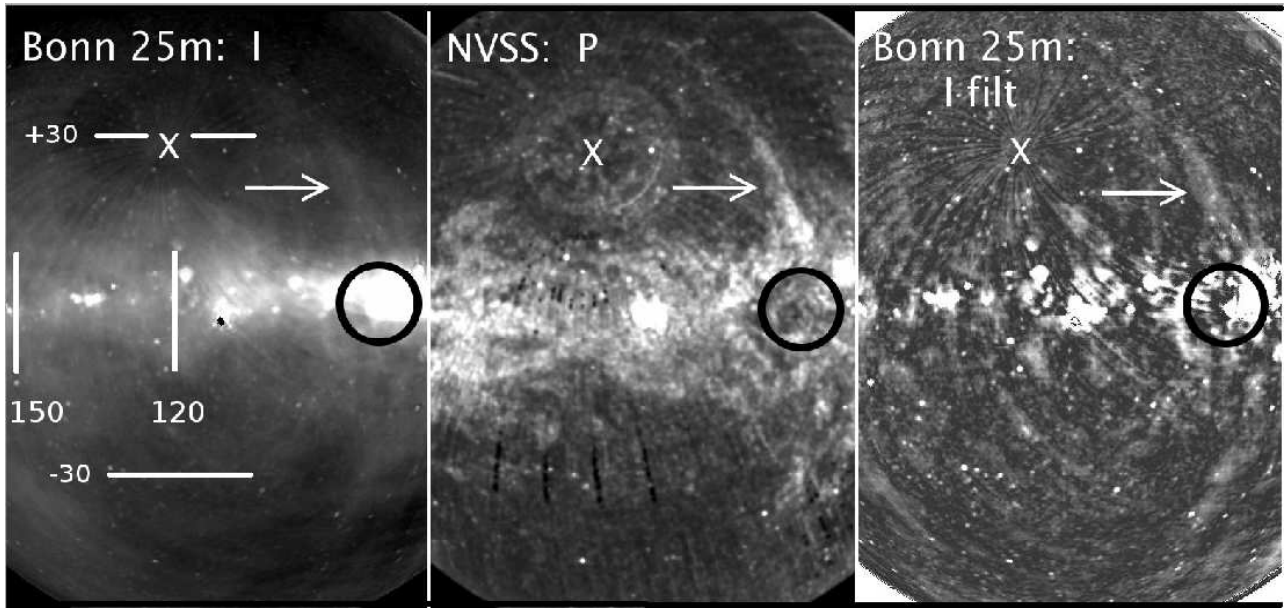}
\end{center}
\caption{Total and polarized intensity around the Perseus-Cygnus region of the galactic plane. Galactic coordinates are indicated on the left image.  "X" marks the position of the celestial pole; circles around the pole  visible in the middle image show lines of constant declination.   Left: 21 cm total intensity from Bonn 25m survey \cite{bonn}, 36' resolution. Center: Polarized intensity from NVSS at 800" resolution, with no filtering.  Right: Filtered version of map on left, as described in text. The black circle marks the Cygnus arm at l$_{II}$=90$^o$; the arrow indicates the base of what is normally called ``Loop III'' and is discussed further in the text and in Figure \ref{newloop}.}
\label{plane}
\end{figure}

 We briefly look at what types of features and artifacts are visible in Figure \ref{allsky} and the additional closer look in Figure \ref{plane} (center image) at a $\sim$100$^o$ region centered around the Perseus arm.  First, we see that the polarization images trace the large scale emission of the Galaxy, as seen, e.g., in \cite{bonn} or \cite{408MHz}.  Superposed on the galactic structure are a network of stripes following lines of constant right ascension.  These lines represent series of scans from the NVSS survey where the residual instrumental polarization was slightly higher than average.  There are also lines of black 'dots', each of them one approximately one primary beam (34') across, where the residual instrumental polarizations were high enough to require flagging in the NVSS survey (W. Cotton, private communication, 2008).  At the highest declinations, the residual instrumental polarization is high at all right ascensions, creating a series of rings around the north celestial pole.

A large number of small sources (at 800" resolution) are also seen in these images.  These represent both real polarized sources and those caused by residual instrumental polarizations from the very brightest objects in total intensity.  To illustrate the instrumental issues from strong point sources, we consider the case of 3C84;  it has an NVSS flux of 22.15~Jy, and an observed polarized flux of 1.9~mJy (0.08\%), showing that the residual instrumental polarization is quite low at full resolution.  However, larger sources such as Cassiopeia~A ($\sim$5' diameter) seen in the center of Figure \ref{plane} are incompletely sampled in the NVSS, producing sidelobe structure that puts power into the polarization images.  Out of Cas~A's total flux of $\sim$2500~Jy, the total polarized flux in  Figure \ref{plane} is $\sim$45~Jy, or $\sim$2\%, spread over $\sim$1$^o$.  Although there may be some small real integrated polarization at 1.4 GHz \citep[see][]{caspol}, the 1$^o$ extent shows that the bulk of the contribution here is instrumental.  Thus, very strong, extended sources cannot be reliably studied through this NVSS reprocessing. We consider this effect further below in our discussion of emission around 3C31. 

By contrast with Cas~A, there appears to be little or no effect due to the strong total intensity emission from the Galaxy, e.g., on scales of degrees or larger. This potentially could have been a problem, as \cite{vgps} show that the noise in VLA visibility data is related to the value of T$_{sys}$, which includes contributions both from the sky brightness and spillover radiation from the ground. However, even in the brightest regions of the galactic plane, we do not appear to have preferentially high polarization;  the brightest region in Figure \ref{plane}, for example, is in the direction of the Cygnus Arm, and is indicated by a black circle.  Within that region, the peak (mean) brightness is $\sim$35~K ($\sim$7~K) from the \cite{bonn} map, while we observe a mean polarized brightness of $\sim$20~mK above the background.  Instead of being higher in this bright region, the NVSS polarized flux is strongly \emph{anti-correlated} with the total intensity emission, likely due to depolarization from the extensive ionized gas seen over this same area \citep{WHAM}.  \cite{stil} also find a major drop in the number of polarized NVSS sources in this area, likely due to depolarization between the two NVSS bands.

A subtle problem with NVSS polarizations has been described by \cite{battye}, largely resulting from clean biases and small ($\mu$Jy) offsets in Q and U. Both of these create problems far below the noise in the NVSS polarization images, and even further below the residual instrumental polarization artifacts that are prominent in Figures \ref{allsky} and \ref{plane}, and do not affect the current work.

\section{Map Analysis}

The key element of our processing is the convolution to larger angular scales of the {\it polarized intensity} maps, instead of the typical procedure of pre-convolving Q and U, and then calculating P. Such \emph{pre-convolution} is
generally preferred due to its better noise properties, as well as its preservation of the polarization angle ($\chi= 0.5 \times tan^{-1}(Q/U)$ ) on the convolved scales.  Pre-convolution is also equivalent to observations made with a large synthesized beam or single-dish observation.  However, since there is a maximum angular scale to which interferometers are sensitive, pre-convolution at that maximum scale or beyond would contain no real signal, only noise.

\begin{figure}[]
\begin{center}
\includegraphics[width=15cm]{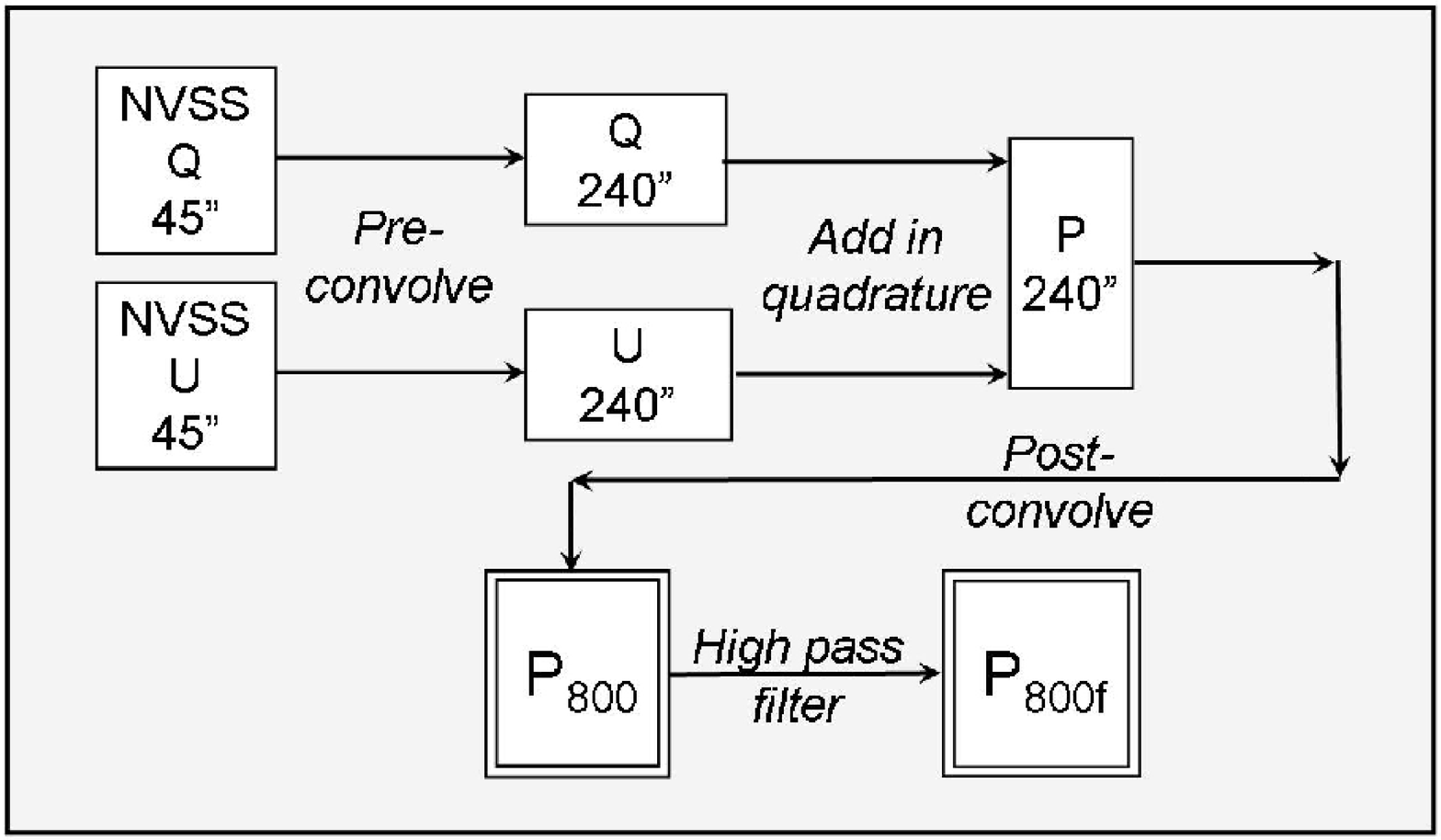}
\end{center}
\caption{Flow chart for construction of polarization images P$_{800}$ and P$_{800f}$. }
\label{pscheme}
\end{figure}

 Once P is calculated from Q and U images at some resolution, further smoothing (which we term \emph{post-convolution} here) can be performed and arbitrarily large structures can be detected. Similarly, if a large total intensity structure were not smooth, but made up of a collection of small-scale features, it would also be detectable beyond the nominal maximum angular scale. This process is diagrammed in Figure \ref{pscheme}.  It shows the production of two maps, \apol and \apolf, constructed as described above by pre-convolution followed by post-convolution at a resolution of 800".  

The rms scatter in \apol and \apolf  varies widely across the sky, and contains contributions from three factors - the random noise from the receivers, the actual polarized signal in each region, and the residual instrumental polarization which is most easily evident in the strong declination stripes visible in Figure \ref{allsky}. During the NVSS observations, the focus was on the reliability of the total intensities, and changes in the instrumental polarization were not closely monitored (B. Cotton, private communication). At times when certain declination strips were being observed, larger than normal instrumental effects would arise, e.g.,  from phase jumps between the right and left hand receivers on the reference antenna used for polarization calibration. We illustrate more quantitatively in Figure \ref{noiselong} how the rms fluctuations in \apol vary across the sky at high galactic latitudes.  We identify the high rms scatter region associated with the actual signal related to the North Polar Spur, and an adjacent region where the residual instrumental polarization contributions were small.  These can also be seen visually in Figure \ref{allsky}.  In the presence of random noise alone, we  could have corrected for the bias in polarized intensity \citep{ward74,sim85}. Given the dominance of variable residual instrumental polarization contributions, however, it is not possible to conduct an automated processing scheme for separating the signal/noise/instrumental contributions to the power.Therefore, we look at the noise characteristics for each individual source of interest, and describe the observed value in \apol  above the local background and rms to estimate its significance.
 
 To remove some instrumental artifacts (and a first order subtraction of the polarization bias) we will also be making use of a high-pass filtered version of \apol, which we label \apolf.  The high-pass filter consisted of subtracting from each pixel the median of \apol over a box 4$^o$ in declination (the size of an individual NVSS image field) by 6' in right ascension.

\begin{figure}[]
\begin{center}
\includegraphics[width=10cm]{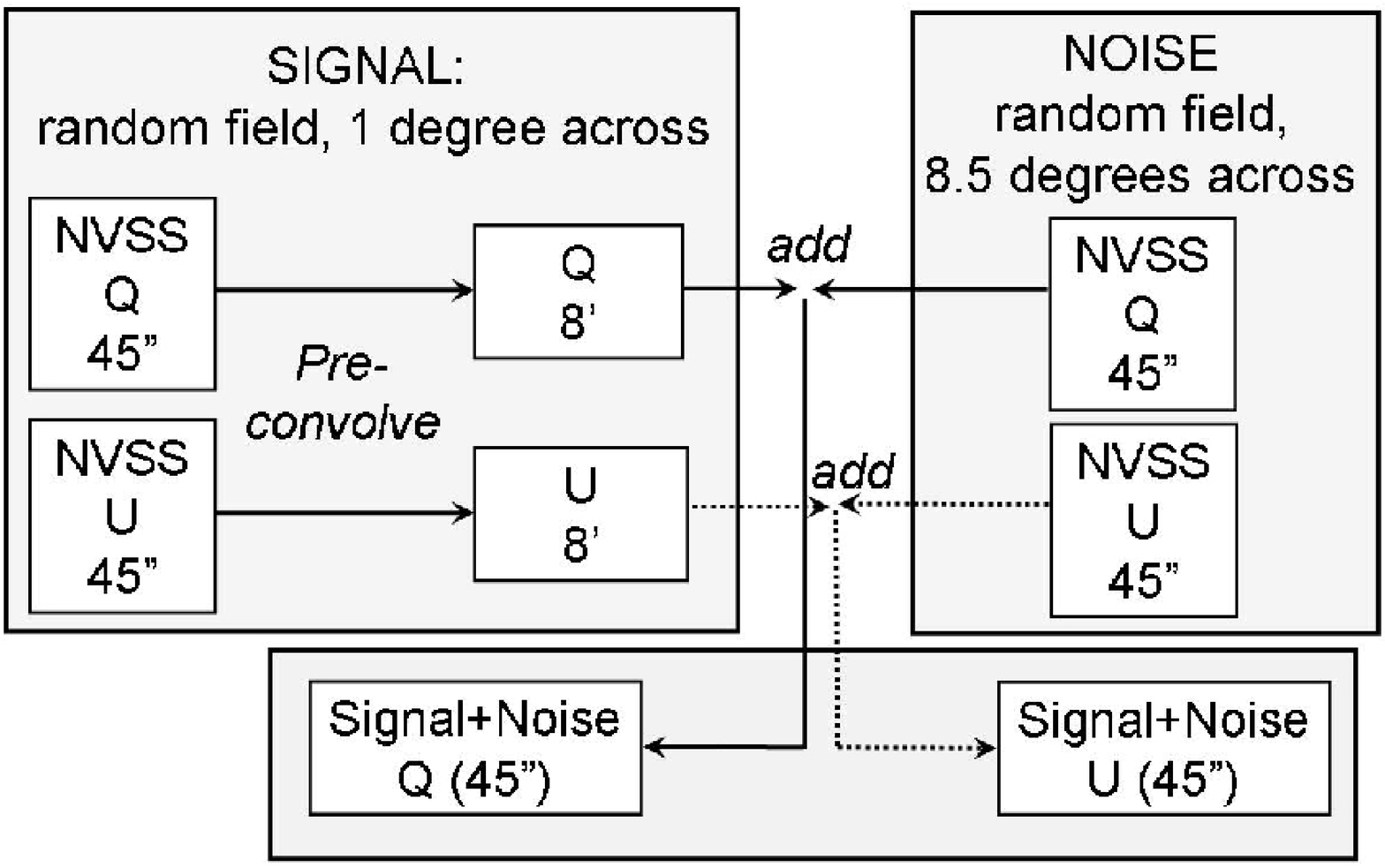}
\end{center}
\caption{Flow chart for construction of simulated images (signal+noise) for sensitivity experiments. The NVSS Q and U images were first clipped to remove the effects of strong, compact polarized sources.}
\label{SNscheme}
\end{figure}

\begin{figure}[]
\begin{center}
\includegraphics[width=10cm]{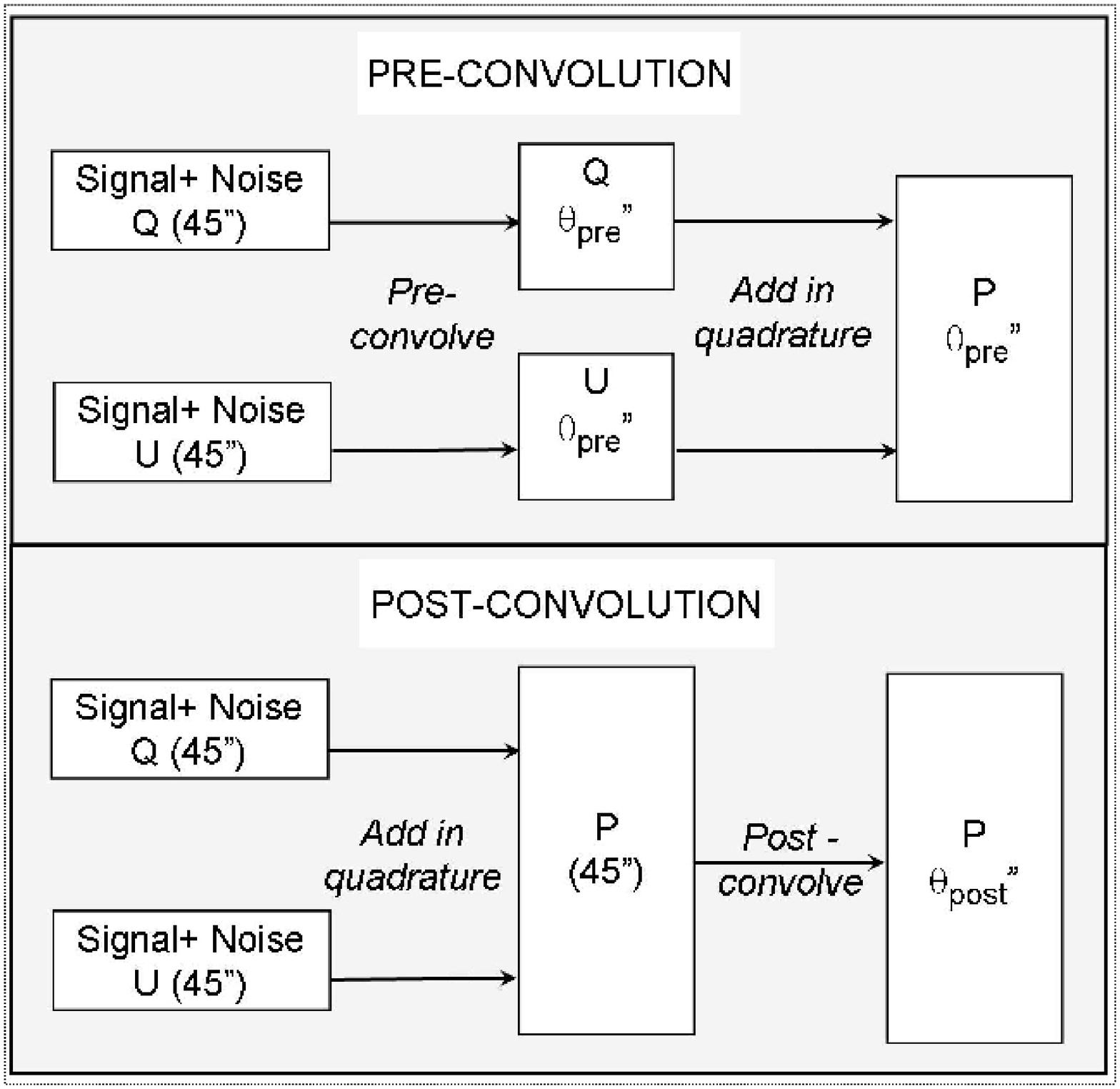}
\end{center}
\caption{Flow chart for two options of processing (Signal+Noise) images for sensitivity calculations.}
\label{SNoptions}
\end{figure}

\begin{figure}[]
\begin{center}
\includegraphics[width=7cm]{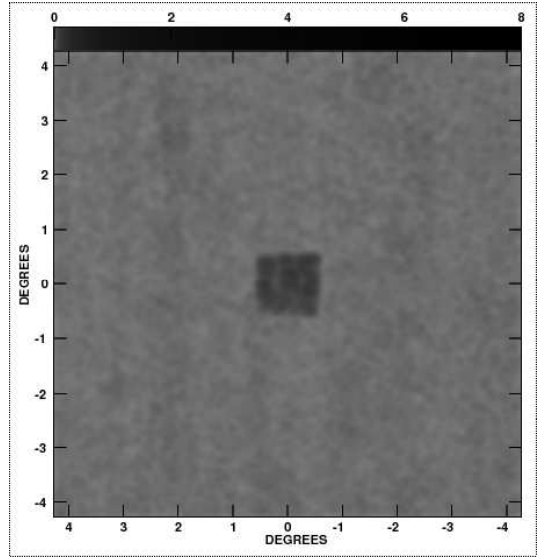}
\end{center}
\caption{Example of simulated polarization data used to calculate sensitivity. The central 1 square degree contains the simulated signal, a random pattern of polarized flux with a characteristic scale of 8'.  In this particular example, the image was ``pre-convolved" by 4' and ``post-convolved" by 8'.}
\label{fakepol}
\end{figure}

\begin{figure}[]
\begin{center}
\includegraphics[width=7cm]{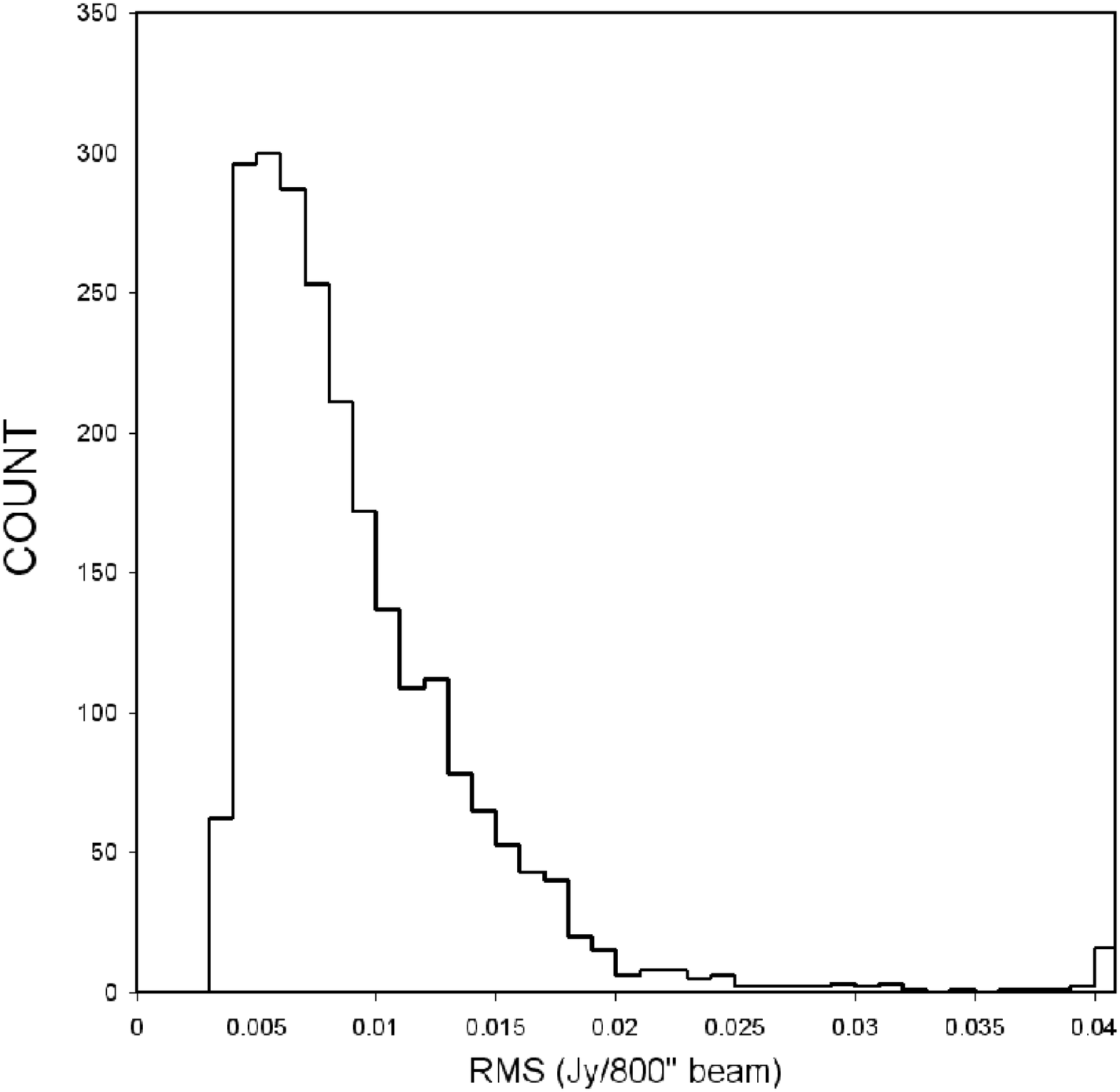}
\includegraphics[width=7cm]{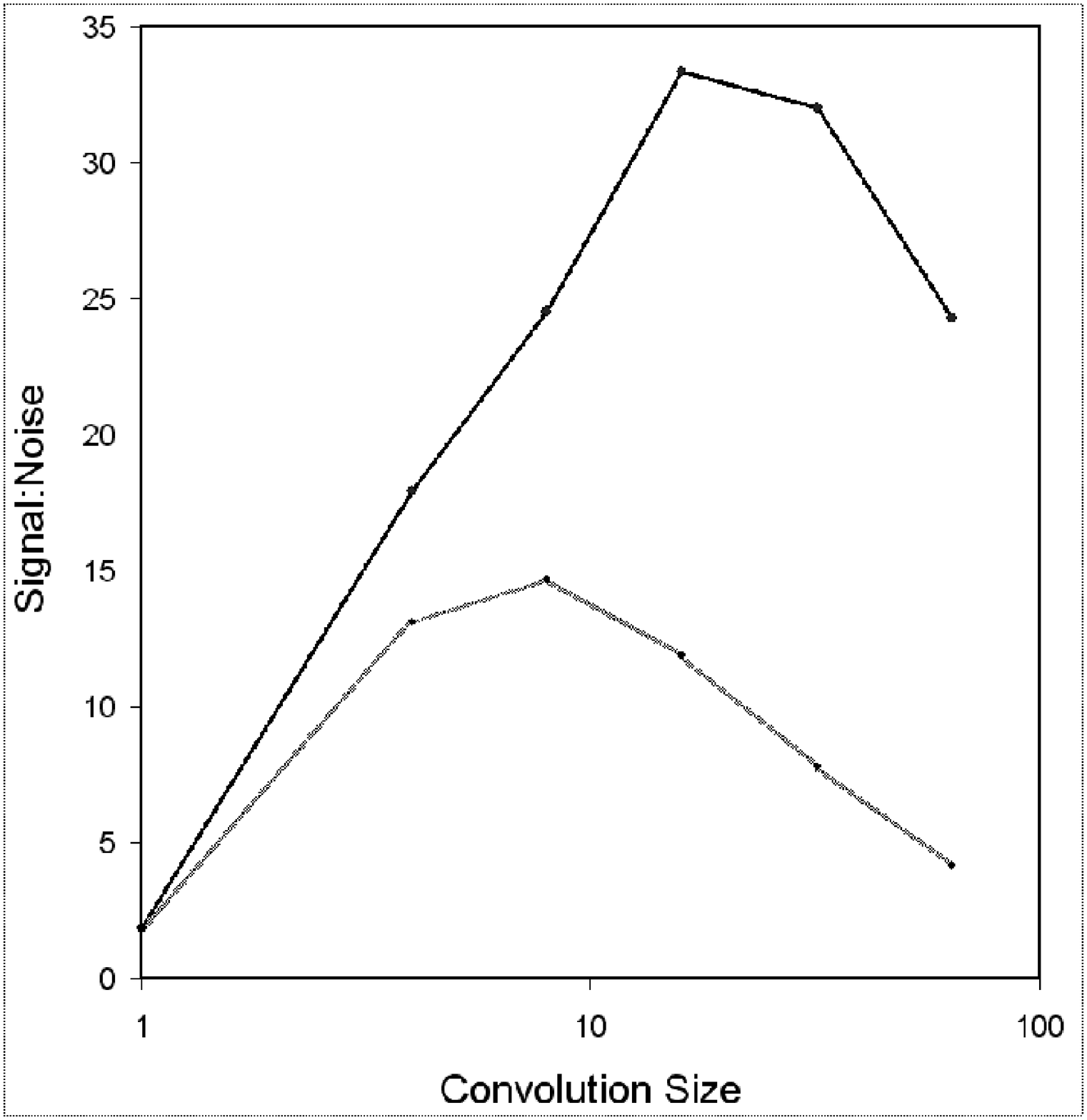}
\end{center}
\caption{Left: Histograms of the rms noise in each 4$^o$ field in the NVSS.  Right: signal:noise from simulated polarization measurements, as described in the text.  The black line indicates "pre-convolution" of the Q and U images; the grey (lower) line indicates "post-convolution" of the P image. }
\label{noise}
\end{figure}

\begin{figure}[]
\begin{center}
\includegraphics[width=15cm]{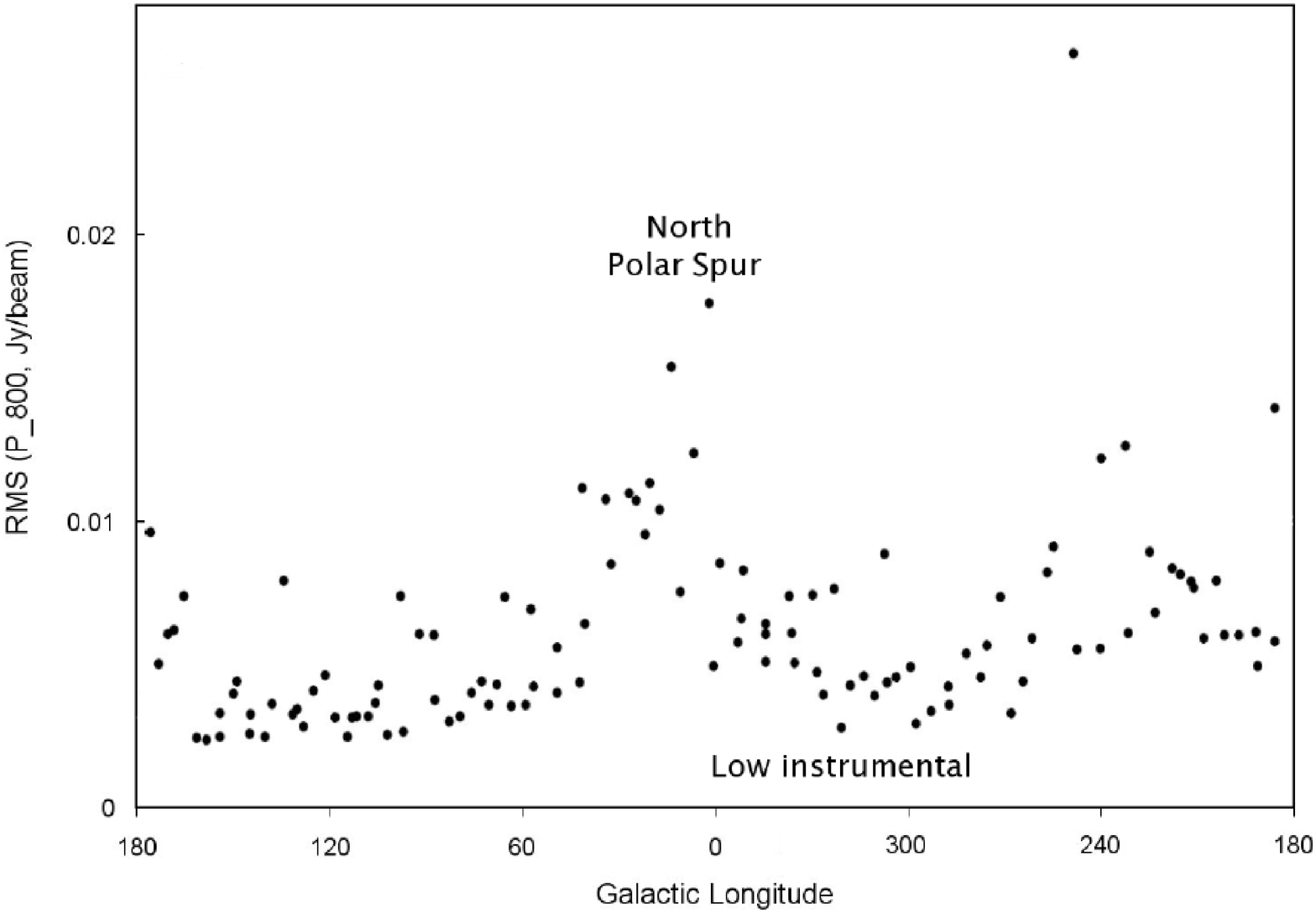}
\end{center}
\caption{RMS scatter in \apol\ in strip from 53$^o < $ b$_{II} < $ 67$^o$ as a function of galactic longitude. Inidividual spikes and the large scatter are due to variations in residual instrumental polarization.  A region of low instrumental effects is indicated, along with the increased rms power due to the North Polar Spur.}
\label{noiselong}
\end{figure}

We estimated the sensitivity to diffuse polarized structures after pre- and post-convolution using a series of simulated polarized signals. The initial construction of the simulated images (Signal+Noise) is shown in Figure \ref{SNscheme}; the processing of these images to measure sensitivities is shown in Figure \ref{SNoptions}. One of the processed (Signal+Noise) images is shown in Figure \ref{fakepol}. In order to capture all the spatial characteristics of the NVSS images, we used actual images from high galactic latitude fields for both the simulated "signal" and "noise".  We constructed the \emph{background ``noise"} image by first taking a single pair of 8.5$^o \times$ 8.5$^o$ Q and U images from the  survey, and at all locations where $\mid$Q$\mid$ and $\mid$U$\mid$ were greater than 2 mJy/beam, replaced these values with zero.  This effectively eliminated almost all signals from compact sources in the Q and U images. The remaining large-scale real and instrumental fluctuations in these images thus simulate the actual background against which we are trying to detect signals. The rms ``noise" distribution, which includes point source, galactic and instrumental fluctuations, is shown in Figure \ref{noise}, for all 4$^o$ fields in the NVSS.  It can be very roughly described as a Rayleigh distribution with an rms value of $\sim$6.5 mJy/(800" beam), slightly larger than the expected system noise value of 5.2 mJy/(800" beam) = 0.29 mJy/(45" beam) $\times ( [800"/45"]^{0.5})$.  At the full resolution of the NVSS,  the systematic real and instrumental effects contribute little power to the Q and U maps.  The systematic contributions become much more important as the images are convolved to lower resolution.
 
 To create our simulated {\emph signal}, we took a separate NVSS 1$^o$ field and convolved the Q and U images with an 8' Gaussian, multiplied these images by a constant and added them respectively to the background noise  Q and U images.  The simulated signal thus consists of a 1$^o$ patch with structures in Q and U on the scale of 8', added to the 8.5$^o$  background (noise) field at the full NVSS resolution of 45". 

We then pre-convolved these images to various scales, converted to P (first option in Figure \ref{SNoptions}) and measured the means in the central 1$^o$ region ($P_{sig}$, including the local noise) and outside the central 1$^o$ (P$_{noise}$). We also measured the rms fluctuations, $P_{rms}$ in the noise regions. Finally, we calculate $$Signal:Noise \equiv [ P_{sig} - P_{noise}]/ P_{rms} .$$  The resulting signal:noise as a function of pre-convolution size is shown in Figure \ref{noise}.  The detectability rises dramatically with convolution, as expected, since the "signal" was forced to have a scale of 8'. At larger pre-convolution values, the signal:noise falls as the signal itself starts to be vector-averaged away and diluted.   

We also show the behavior of the signal:noise using \emph{post-convolution}. For these experiments, we added the ``signal" image (with its 8' scale) to the full resolution Q and U ``noise" images, formed P, and then post-convolved the result to various scales (second option in Figure \ref{SNoptions}).  Looking again at the signal:noise for various convolutions, we see that the detectability of the signal is less than in the vector-averaged (pre-convolution) case because the signals are not being averaged coherently. At the larger convolution sizes, the signal:noise drops even further, mostly due to the increased noise power on large scales in the NVSS, from both galactic polarization structure and residual instrumental polarization.   We note that the above signal detectability experiment is a single case using an arbitrarily constructed source.  The detectability of sources with specific polarization structures would have to be investigated on an individual basis.

It is also possible adopt a hybrid approach and pre-convolve an image to some scale ($\theta_{pre}$), and then post-convolve it to larger scales ($\theta_{post}$).  The advantage to this approach is that if an astronomical source has structure in Q and U on scales $\sim\theta_{source}$, then the signal:noise will be enhanced by using $\theta_{pre} \sim \theta_{source}$.  This pre-convolution has a second advantage for our survey -- it enhances our search for new larger-scale features by enhancing them relative to smaller sources that are more likely to be already known.   We therefore adopt, for this initial presentation $\theta_{pre}=240"$ and $\theta_{post}=800"$, and call the polarized intensity \apolf . It is important to note that while our \apolf maps with $\theta_{post}=800"$ have a resolution very similar to those of single-dish surveys, sources with $\theta_{source} \sim 240"$ will be detected by us, but {\it not} with the single dish.

The results presented here are almost entirely qualitative, for a number of reasons.  First, they measure only the polarized ``power''in Q and U from the poorly sampled low spatial frequencies in the survey up to  the pre-convolution scale size.  Thus, a source one degree across with uniform Q,U would be invisible, while another source with variations on, e.g.,  4 arcminute scales would appear strongly, even if they had the same mean $sqrt(Q^2 + U^2)$.  The spatially dependent power in Q and U may also depend on the foreground rotation measure, independent of the intrinsic strength of the source.  Quantitative information could be derived only by constructing a detailed polarization model of a source and its foreground rotation measure structure, and then propogating that model through simulated observations and processing.  We do not attempt this here.

\section{Results}

We begin with a discussion of two large fields in the galactic plane, to illustrate the differences between structures seen in total intensity and/or single-dish polarization maps.  We also look at several new diffuse polarized structures, galactic and extragalactic, found through our new processing.  A paper identifying a large number of new and often unidentified sources is in preparation.

\subsection{Cygnus-Perseus region}
\begin{figure}[]
 \begin{center}
  \includegraphics[width=15cm]{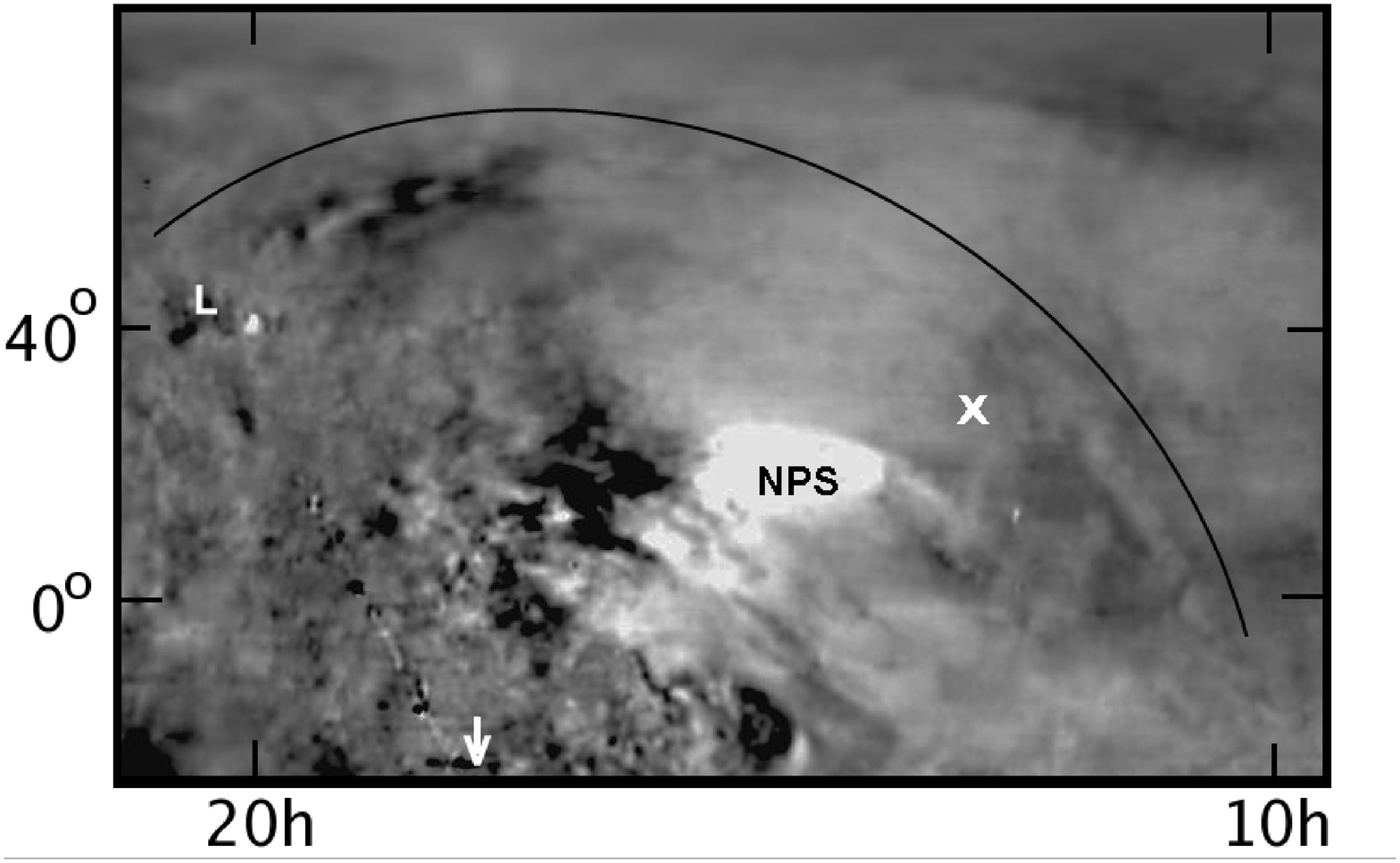}
 \end{center}
\caption{DRAO Stokes Q image \citep{DRAO} in celestial coordinates, centered at RA,, Dec. $\sim$ 15h, 30$^o$ , and approximately 150 $^o$ in RA by 110$^o$ in Dec. ''L`` marks the base in the galactic plane of the suggested new loop, at l$_{II}$=82$^o$. while the arrow shows the position of the galactic center, just below the edge of the image, so the galactic plane runs between these two.  X marks the galactic pole, and NPS denotes the North Polar Spur. A curved line is drawn outside of the possible new loop.}
\label{newloop}
\end{figure}

This region is shown in Figure \ref{plane} and we have previously discussed some of its instrumental and other features.  Here, we point out the  filamentary feature extending north for about 40$^o$ from the galactic plane in the Cygnus region (l$_{II}$=90$^o$) with a less-well defined counterpart to the south.  The filament is also present in the total intensity Bonn image at 36' resolution \cite{bonn}, especially when filtered to remove the smooth background (using the multi-resolution filtering technique of \cite{rudfilt} with a box size of 7$^o$ in longitude by 1$^o$ in latitude).  This feature is usually identified as the western portion of \emph {Loop III} \citep{loop3}, but another possibility is suggested in Figure \ref{newloop}.   The 36' resolution Stokes Q image from \cite{DRAO} is shown here, with the suggestion of a new loop that extends from the Cygnus region, curves up and over the galactic pole, and returns back towards the galactic plane approximately 180$^o$ away.  With a nominal center at l$_{II}$=0$^o$, b$_{II}$=20$^o$, and a radius of 75$^o$ (M. Wolleben, private communication), it is approximately concentric with Loop I, which  \cite{woll07} has recently modeled in terms of two $\sim$100~pc radius synchrotron emitting shells  in which we are embedded. This new suggested loop needs further study through other ISM tracers, and could potentially revise our understanding of our local environment.

\subsection{Galactic Center region}

\begin{figure}[]
\begin{center}
\includegraphics[width=15cm]{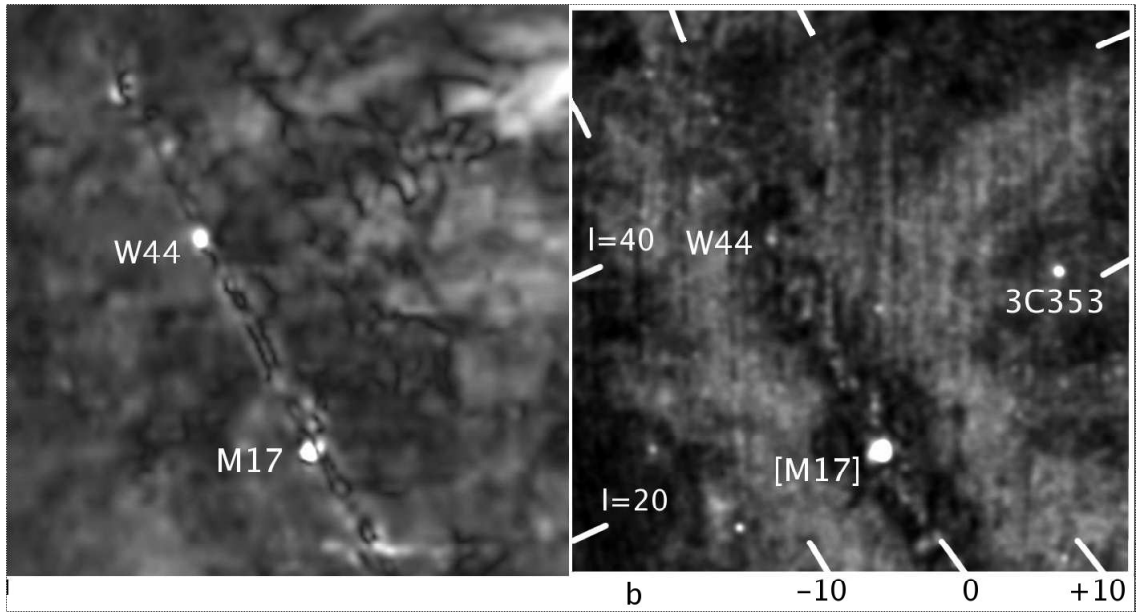}
\end{center}
\caption{Left: Polarized intensity from DRAO survey \citep{DRAO}. Right: Polarized intensity from NVSS.  Both images have a resolution of 36'.  The field is centered at l,b =  26.5$^o$, -1.1$^o$ and is 50 degrees in width.}
\label{nvssdrao}
\end{figure}

A 50$^o$ field near the galactic center (Figure \ref{nvssdrao}) further illustrates the similarities and differences between single dish polarization measurements and the NVSS polarization reprocessing at the same wavelength and angular resolution.  We pre-convolved the NVSS Q, U images with 240" and post-convolved the P images with 36'. In the NVSS image, the vertical bands are artifacts from small differences in the residual instrumental polarization at different times, as observations were made along lines of constant right ascension.  This is also likely responsible for some of the fine-scale mottling apparent along the vertical bands, but it is not possible for us to separate these from fine scale polarization structures.  We believe that the rest of the structures visible in this NVSS image are ``real'' in the sense that they reflect the actual signals from the sky.  Their interpretation, however, is quite different from those of single dish images, as described below.

 As is well known, polarization and total intensity structures are very different, even when they are both from single-dish measurements.   The image from the  DRAO polarization survey \citep{DRAO} shows a bright patch in the upper right, with a smoothly varying position angle (not shown) over scales of degrees.  The NVSS is not sensitive to Q and U variations on such large angular scales, and the feature is not visible in our images.  High rotation measures ($>$340~rad/m$^2$) would also cause a degradation of the NVSS signal compared to the DRAO image.  Along the galactic plane, running from upper left to lower right, the DRAO image shows an unresolved narrow bright band bordered by two dark stripes.  The dark stripes are beam-depolarized due to a rapid switch in polarization angle across these lines.  In the NVSS data, these same patterns in the sky create a different response;  a broken thin strip of polarized emission along the plane  is seen from the regions where the angle is changing rapidly, creating a detectable interferometer signal in Q and U.  However, farther from the plane,  the relatively constant polarization angles from the DRAO position angle image (not shown here) lead to a broad dark region in the NVSS image.  The numerous depolarization filaments seen elsewhere in the DRAO image are again not seen in the NVSS;  rapid changes in polarization angle provide an NVSS signal at high resolution, and there is no depolarization introduced by the post-convolution.  An alternative probe of depolarizing regions is presented by \cite{stil}, who study the polarization of compact sources in the NVSS.  They find regions $\sim$10$^o$ across that cause a reduction in the number of polarized NVSS sources, due to both intervening H~II regions and diffuse galactic structures.

In Figure \ref{nvssdrao}, we also point out the bright emission from 3C353 in the NVSS image. This radio galaxy has very strong polarized (Q and U) emission when convolved to the 240" scale, which is then carried forward through the post-convolution to 36'.  However, the averaging of Q and U themselves over 36' in the DRAO image makes 3C353 undetectable against the galactic background.

The very luminous HII region M17 is bright in both the DRAO and NVSS images.  The NVSS apparent extended polarized flux comes from strong sidelobes, as can be seen in the full resolution images; it is likely that the DRAO polarization is also due to instrumental polarization.  W44 is detected strongly in DRAO and weakly in NVSS.  The full resolution NVSS polarization image (not shown here) shows a structure which is largely unrelated to the total intensity structure of W44, so the physical origin of the polarization seen at 36' resolution is unclear.

\subsection{The Extragalactic Sky}

\begin{figure}[]
\begin{center}
\includegraphics[width=19cm, angle=90]{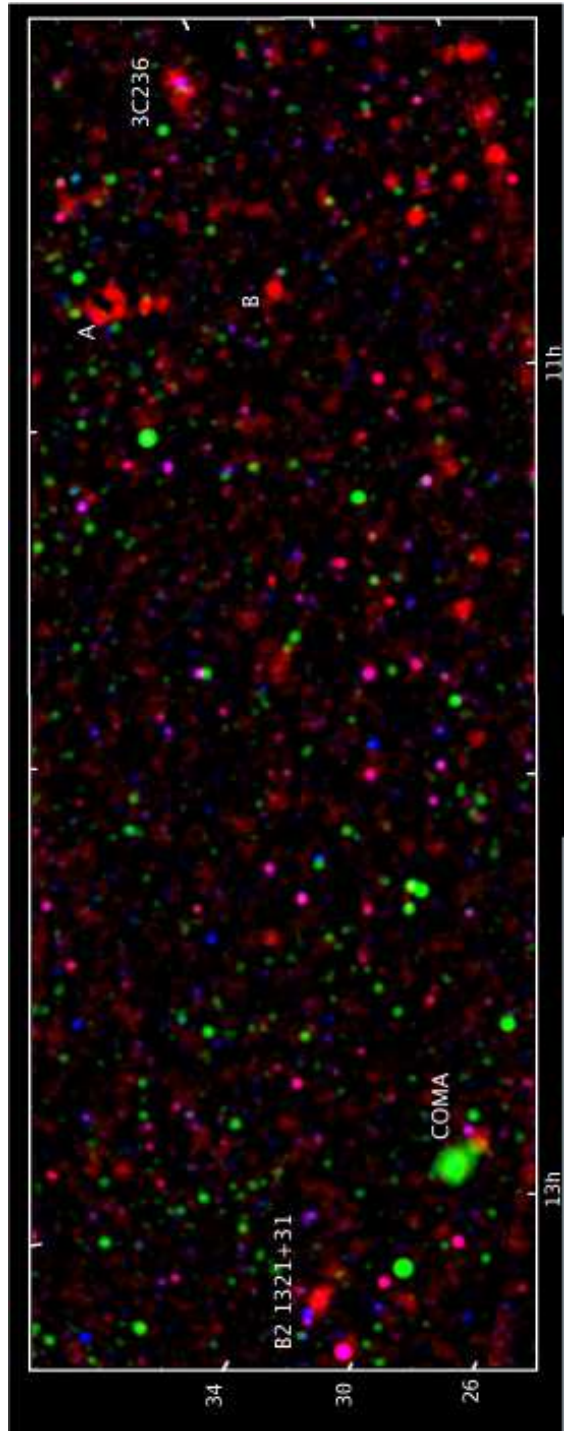}
\end{center}
\caption{Image of approximately 44$^o\times$17$^o$ strip in celestial coordinates, centered around 11.8h,34.2$^o$  Red is \apolf, green shows the broadband X-ray emission from ROSAT, convolved to 800", and blue is the total intensity emission from NVSS, convolved to 800".}
\label{RGB}
\end{figure}

Away from the galactic plane, we get a better view of more compact polarization features in the NVSS, as seen in Figure \ref{RGB}. This is an $\sim$~44$^o \times$ 17$^o$ strip of the sky  centered around RA, Dec 11.8h, 34.2$^o$, with \apolf in red.  The green image shows the ROSAT All-Sky Survey \citep{rass} convolved to 800".  The blue image is the total intensity NVSS image, also convolved to 800".   Note that at this high galactic latitude ($35^o~<~ b_{II}~<~90^o$), and with the additional filtering, very little galactic structure is visible in \apolf.  A large number of bright compact features are visible, appearing in red (blue) for high (low) fractional polarization.  Most of these are also seen in the full resolution NVSS images. 

A number of more extended polarization features are also visible in Figure \ref{RGB}, many of which are well-known sources.   In the southeast, the bright green structure is the X-ray emission from the Coma cluster of galaxies, with an extension to the southwest from an infalling subcluster \citep{ferr06}.  At the southwest X-ray terminus is a transverse polarized radio structure with brightness $\sim$20~mJy/800" beam (20 mK) above the background. This is the well-studied ``relic" first detected by \cite{jaffe79}, and then studied in total intensity and polarization, e.g., by \cite{hanisch,giov91}.   Note the absence of blue, total intensity, emission from this polarized structure; with an angular extent of $\sim$1$^o$, the total intensity is not detected in the NVSS survey.   Similarly, the diffuse polarized emission from the giant radio galaxies 3C236 and B2~1321+31 is easily visible, while only their more compact features are seen in total intensity in the NVSS.

Among the bright extended polarization structures are several that were previously unknown.  Feature A is one of the brightest structures in the high latitude sky (25 - 35 mK); follow-up observations with the Westerbork Synthesis Radio Telescope \citep{brown07b} suggest that it is a galactic Faraday system.  Such structures have no intrinsic synchrotron emissivity, but appear bright to the interferometer because they produce small-scale variations in Q and U from the galactic background \citep{wollfar}.  Feature B, and others like it in the image, are examples of single NVSS pointings with high residual instrumental polarization; they are recognizable by their circular shape and sizes comparable to the primary beam ($\sim$30').  We have not yet found a reliable way to eliminate instrumental problems at the lowest levels seen in Figure \ref{RGB}, especially when they span several pointings.  In the following, we therefore present examples of new extended polarization features only with high signal:noise, to illustrate the potential power of this NVSS reprocessing technique.

 \subsection{Abell 3744}
\begin{figure}[]
\begin{center}
\includegraphics[width=9cm,angle=-90]{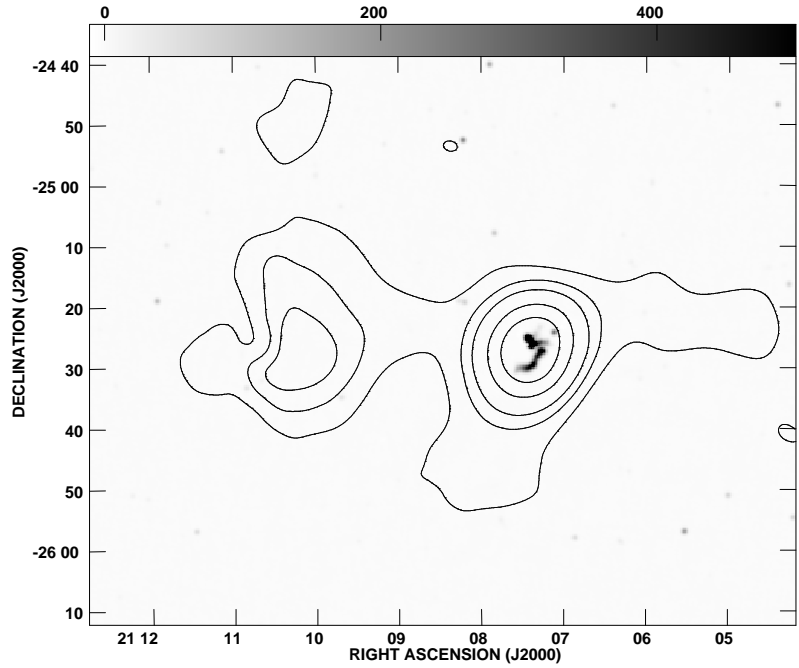}
\vskip .5in
\includegraphics[width=9cm,angle=-90]{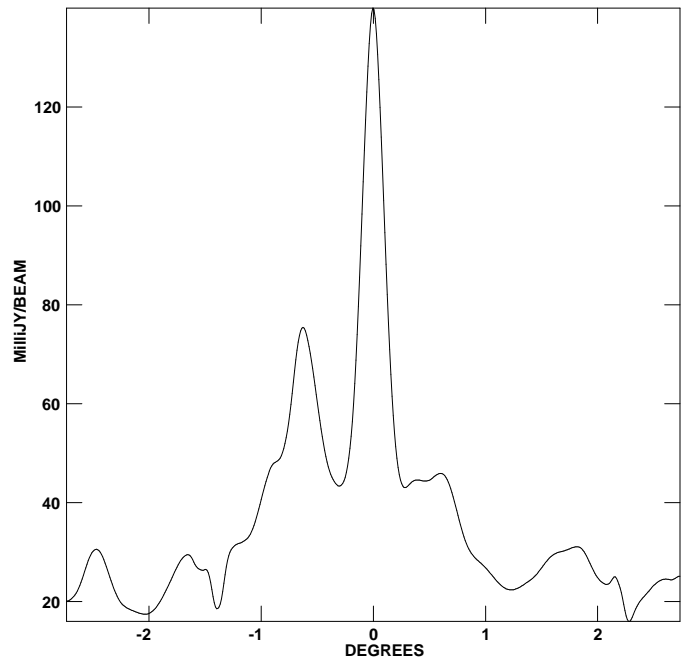}
\end{center}
\caption{Abell 3744, contours and slice of NVSS polarized intensity at resolution of 800".  The top image is overlaid by the NVSS total intensity at 45" resolution. Contour levels are at 0.01 $\times$ (4, 5, 6, 8, 10) Jy/beam.  The total field is 95" in declination by 115" in right ascension. The slice is taken in right ascension through the peak and extends over 5$^o$ to show the signal:noise of the detected features.}
\label{a3744}
\end{figure}

 A network of tailed radio sources is seen in this cluster \citep{marvel}, which might be responsible for the bright polarized emission centered on the cluster seen in Figure \ref{a3744}.  At a redshift of 0.0381, Abell 3744 is listed as a member of supercluster number 180 by \cite{einasto}, who group it with Abell 3733, (z=0.0382), and, apparently mistakenly, with the background cluster Abell 3706 at a redshift of $\sim$0.1 .  Abell 3744 is a member of the REFLEX sample \citep{REFLEX} with a relatively low X-ray luminosity of 1.6$\times$10$^{43}$ erg/s, integrated over 0.1 - 2.4 keV. The X-ray emission is seen only in the region of the cluster center.
 
 To the east, a polarized structure is seen 37' (1.8~Mpc) from the cluster center, with an extent of $\sim$1.4~Mpc. A slice at constant declination through the peak of polarized emission and through the eastern structure is also shown in Figure \ref{a3744}, to demonstrate the signal:noise of these features.  The eastern structure has a peak polarized flux 40 mJy/beam above the background, with a background rms of 6 mJy/beam.  It has a total polarized flux (\apolf) of $\sim$90 mJy, and no obvious total intensity counterpart. It is important to note that our polarized fluxes are the result of our non-standard processing procedure, and that the true signals measured with an interferometer plus single-dish system could be considerably higher, assuming no additional contributions from a nearby Faraday screen.   Assuming our nominal fluxes and a 33\% fractional polarization would yield a monochromatic radio luminosity of 10$^{24}$ W/Hz at 1.4 GHz.  If this is a peripheral relic, or gischt \citep{kempner}, both its radio and associated X-ray cluster luminosities are significantly lower than those of most relic systems \citep[e.g.,][]{giov99}.  Only one other cluster, Abell 548b, which is part of a very complex optical and X-ray system \citep{davis}, has similarly low luminosities.  The Abell 3744 relic would also be considerably further from the cluster core than is typically seen.  Confirming and more detailed polarization observations of this system would also be most useful.
 
\subsection{3C 31}%user 118, faraday cat. 291}
\begin{figure}[]
\begin{center}
\includegraphics[width=10cm]{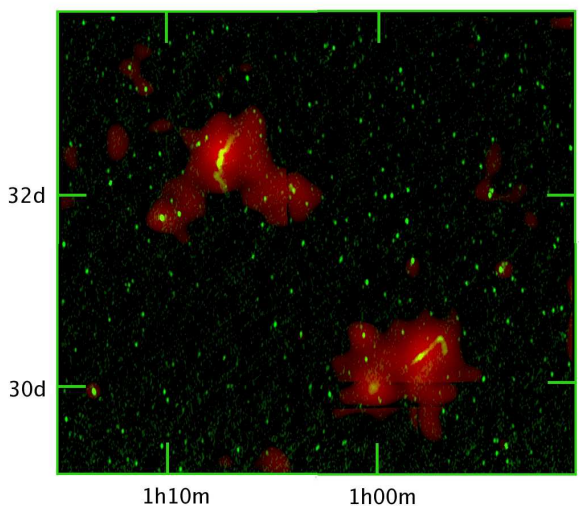}
\end{center}
\caption{5.3$^o$ field with 3C31 (upper left) and NGC~315 (lower right). \apolf is in red.  The brightness of the southwest extension of 3C31, ignoring the region of the background double, is $\sim$12~mJy/800" beam. The vertical and horizontal lines indicate slight differences in background removal from the original 4$^o$ NVSS fields. The green image is the full resolution WENSS survey image at 330 MHz.}
\label{3C31}
\end{figure}

3C31 is a very well-studied wide-angle tailed radio galaxy \citep{fanaroff} with a total extent of $>$40' \citep[$\sim$850 kpc at z=0.0169 with H$_o$= 70 km/s/Mpc,][]{klein}.  In our polarization image, Figure \ref{3C31}, we find emission extending $\sim$1.2$^o$ to the southwest, where it partially merges with polarized emission from an unrelated background source.  To the southeast, there is polarized emission bridging to 3C34, an unrelated source in a z=0.69 cluster \citep{mccarth}. 
Formally, the statistical signifance of the polarized emission is high.  Towards the southwest (southeast), the polarized brightness is $\sim$15 (25) mJy/800'' beam above the background, with a background rms of 4~mJy/800'' beam.  However, there is an additional source of contaminating polarized emission in the neighborhood of strong, very extended polarized sources such as 3C31.  These are sidelobe structures from the poorly sampled short baselines, but would appear as true positive-definite signals in our processing.   To assess the importance of this effect near 3C31, we constructed an equivalent `$I _{800}$'' image by processing the total intensity image in an identical way to \apol, substituting I for both Q and U.  This created excess power in the vicinity of 3C31, as expected.  However, the ratio of the brightness in the southwest (southeast) polarized feature to the peak brightness of 3C31 itself is 0.1 (0.17) in \apol, whereas the brightness ratios for the same locations in $I_{800}$ are $\le$0.01.  We therefore conclude that poorly sampled sidelobes from the polarized emission in 3C31 is not responsible for the newly detected features.

\begin{figure}[]
\begin{center}
\includegraphics[width=8cm]{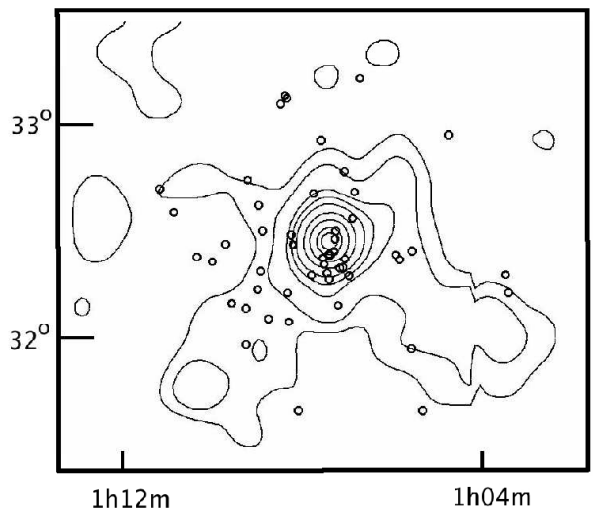}
\label{3c31gals}
\end{center}
\caption{Contours of polarized emission (\apolf) in a 2$^o$ field around 3C31, at levels of (6, 13, 28, 38, 51, 76, 100, 126 and 140) mJy/800" beam.  Small circles indicate the positions of galaxies at redshifts of 0.014 to 0.020 from the 2MASS survey \citep{2massz}.}
\end{figure}

  It is not clear whether this diffuse emission is associated with 3C31 or 3C34. The host galaxy of 3C31 is NGC~383, the brightest of a rich group of galaxies \citep{burbidge}, many of which are seen in the area of the polarized emission to the SE of the source (Figure \ref{3c31gals}).  These are all part of a filament of galaxies at this redshift \citep{miller},  which extends through the SW polarized extension over 9$^o$ to the NE to cluster Abell~262 \citep{moss}. In the $\lambda$11.1~cm single-dish images of \cite{klein}, a 10' extension is seen to the southwest.   The low resolution 102 MHz images of \cite{artyukh} also show extended emission to the southwest, with a size $\leq$50', the extent of their primary beam. There is possible contamination from the southwest background source.  Note that in their Figure 1, 3C31 is the northern component of the apparent double structure; the southern component is 3C34.  

Assuming that the southwest extension is associated with 3C31, it would have an extent of $\sim$1.5~Mpc.  There is some possibility that the observed polarization structure results from distant sidelobes of 3C31, although processing the total intensity NVSS images in the same way as Q and U (i.e., forcing them to be positive and then post-convolving), did not show significant sidelobe contributions.  Assuming the polarization structure is real, the $\lambda$11.1~cm results cited above also suggest that there has been an outflow to the southwest from 3C~31.  Lacking bright jet features, this structure would likely be a remnant of past activity, perhaps now being energized by large scale group processes. Indications of the dynamical state of the group include the offset of  NGC~383 from the centroid of the X-ray emission and significant X-ray structure towards the southwest \citep{kormossa}, which is detected out to the virial radius at $\sim$700 kpc.

 \subsection{CTA1}

\begin{figure}[]
\begin{center}
\includegraphics[width=14cm]{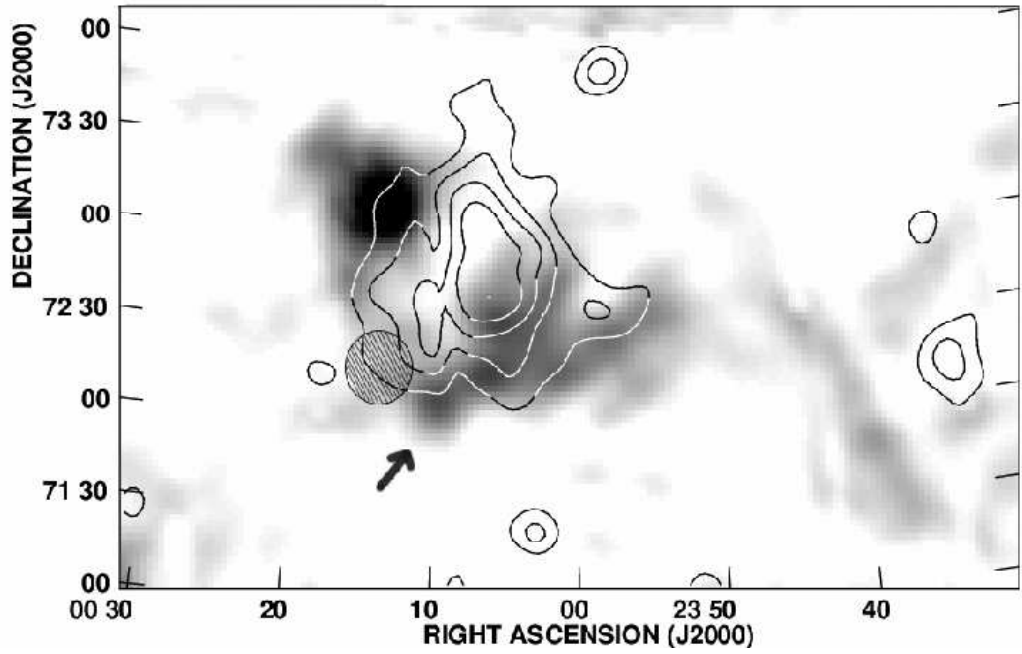}
\end{center}
\caption{CTA 1.  Greyscale is \apolf, with a peak brightness of 45 mJy/800" beam.  Contours are broadband X-rays from ROSAT, convolved to 800".  The hatched circle represents the cloud described by \cite{pin97}, and the arrow indicates the new polarized patch not visible in the DRAO higher resolution image.}
%levs 1,2,3,4,5,6 e-4 cps}
\label{CTA1}
\end{figure}

 This well-known supernova remnant has a filled X-ray structure and a radio shell which is not visible in the NVSS total intensity images.  It has been mapped at 1' in polarization at 1.4 GHz using the DRAO telescope \citep{pin97}.  All of the major features seen in the DRAO image are reproduced in our NVSS reconstruction at 800" resolution (Figure \ref{CTA1}).  In addition, we find one diffuse bright patch at 00h10m, 72$^o$ that is not seen in the DRAO map. The diffuse nature of this patch is established by the lack of smaller-scaled polarized features in the full resolution Q, U images (at a level of Q, U $<$ 3~mJy/45" beam). Its peak brightness is 30 mJy/800'' beam above the local mean, with a background rms of $\sim$5 mJy/800'' beam.   If it were completely smooth, the polarized brightness would be only  0.2~mJy/1' beam, compared to the DRAO rms value of 0.3~mJy/1' beam, so it is reasonable that it had not been previously detected.  %There is still a puzzle with the cloud encounter hypothesis, in that bright synchrotron emission should have been observed at DRAO on the leading edge of the encounter, but this is not seen.
 
 The bright patch is on the border of what \cite{pin97} call the "reverse shell" region, where the curvature of the radio structure is inverted.  Their explanation for the shape of this region is that the shock has encountered and encircled a dense cloud.
 The new polarized patch may help define the borders of this cloud encounter, but the lack of bright, narrow polarized or total intensity emission, as pointed out by \cite{pin97}, is still a problem for this explanation.

%\subsection{Dust}
%There appear to be a number of places where the 100$\mu$m dust emission \citep{dust} is \emph{anti-correlated} with the polarized brightness in our NVSS reprocessed data.  One such region is shown in Figure \ref{dustpol} centered at approximately (l,b)~=~(170$^o$, 2$^o$).  We are unaware of existing work describing such an anti-correlation, and its cause is presently unclear. The low polarization could result from an overall large Faraday rotation, or structure in polarization at very small scales ($<$45") which would depolarize in the NVSS beam. Small-scale structure could arise from dense Faraday-rotating patches in the dusty regions, or  the small-scale magnetic field variations expected in regions with high dust densities, such as described by  \cite{klebe}.  The dust/polarization anti-correlation is also likely related to the strong and well-studied anti-correlations \citep{gaensler} with emission measure and H$\alpha$, although the dust and H$\alpha$ \citep{halpha} structures are not well-correlated in this particular area.

\section{Discussion}

Polarized intensity is not a well-defined quantity when there is more than one polarized component along the same line of sight, creating both challenges and opportunities. Although there are heroic attempts to separate multiple components if they are separated in rotation measure space \citep{brent, schnitz}, in general,  observed polarized intensities depend on a complicated combination of observing frequency, resolution, and the spatial frequencies represented in the image.  This complication also allows a variety of processing schemes to be carried out on Stokes parameter images, which can result in new structures being identified. 

Our reprocessing of the NVSS survey in Stokes Q and U explores a region of observing frequency/resolution/spatial frequency that has not been previously studied.  We therefore have been able to identify many new features, as well as recover structures that are invisible in the total intensity NVSS but seen at other telescopes.  One common feature of the extended emission features we detect is that they are of low surface brightness.  \apolf images free of galactic emission have rms fluctuations $\sim$2~mJy/800" beam (2~ mK).  A 5$\sigma$ detection would therefore correspond to $\sim$10~mJy/800" beam or $\sim$30~mJy/800" beam in total intensity, assuming a fractional polarization of $\frac{1}{3}$.  By contrast, a single dish working at the same frequency and angular resolution, would have a 5$\sigma$ confusion limit of $\sim$125~mJy \citep[e.g.,][]{confuse}.  Under favorable conditions of high fractional polarization and Q,U variations on scales $<$15', the reprocessed NVSS can then be much more sensitive to diffuse structures than any single-dish survey.  

All-sky surveys of diffuse polarization could thus be a powerful way to probe low density extragalactic regions. Assuming an optimum sensitivity of 30mJy/800" beam (total intensity), the equivalent minimum energy magnetic field for a detected synchrotron structure will have values $\sim$0.2$\mu$G. This corresponds to pressures of $\sim$10$^{-14.5}\ \frac{erg}{cm^3}$, in the regime of the WHIM \citep{WHIMkang}.  Radio observations at these low brightness levels could then serve to illuminate the diffuse baryon component of large scale structure, where 30-50\% of the baryons are likely located, but are exceedingly difficult to detect through their thermal emission.  Simulations by \cite{miniati04} and \cite{pfrommer06}, e. g., show that such radio emission is expected, driven by the magnetic field amplification and relativistic particle acceleration at shocks on the borders and in the interior of filaments.

Much more sensitive polarization measurements, with significantly reduced instrumental problems, will soon be available on the EVLA \footnote{http://www.aoc.nrao.edu/evla}, and a new generation of wide-field, low frequency instruments such as LOFAR \footnote{http://www.lofar.org} and the MWA \footnote{http://www.haystack.mit.edu/ast/arrays/mwa} will allow us to probe much deeper into the WHIM regime over large areas of the sky.  Combinations of single dish and interferometer observations may also allow such features to be seen in total intensity for individual regions where the large investment of observation and analysis time can be justified.

\subsection{CMB polarized foregrounds}

It is unlikely that features found through our NVSS reprocessing will affect upcoming CMB polarization experiments. We have identified new features on scales down to $\theta_{pre}$ = 240". For the new synchrotron sources, forthcoming CMB polarization experiments that work at these resolutions or smaller, such as Planck or 
EBEX \citep{oxley04}, \footnote{See http://lambda.gsfc.nasa.gov for a complete list of CMB polarization experiments.} will have sufficient frequency coverage to model the foreground contamination 
from galactic or extra-galactic synchrotron sources internally.  We also detect ``pseudo-sources" due to Faraday screens in our galaxy. However, direct Faraday rotation of the CMB signal from these regions is 
negligible at $\nu > 100$GHz ($\Delta \theta \approx 0.5\deg$ for a rotation measure of RM=1000 rad/m$^{2}$). At the same time, the Faraday screen features could indicate the 
presence of undetected HII regions \citep{sun07}. The bremsstrahlung emission from HII regions can be polarized at the 10$\%$ level due to Thomson scattering at the edges of the clouds 
\cite{keating98}, but the polarized brightness should be at least an order of magnitude less than that of synchrotron emission at frequencies above 10 GHz 
\cite{bennett92}.  Such signals are typically not modeled as a polarized foreground component in the WMAP analysis \citep{page07,gold08}. 

\section{Conclusions}
A reprocessing of the NVSS polarization images has allowed the recovery of diffuse structures on large angular scales.  The details of the processing allow one to tailor the sensitivity to particular angular scales of interest.  At present, residual instrumental polarization variations across the sky are a key limiting factor.  A variety of new galactic and extragalactic sources have already been identified, with a more comprehensive census underway. While a better recovery of diffuse polarization structures is possible, e.g., by combining single dish and interferometer measurements, the next generation of radio telescopes such as LOFAR, the MWA and EVLA will also be able to exploit the processing technique introduced here to provide probes, e.g.,  of the relativistic plasmas associated with the elusive WHIM.

\begin{acknowledgements}
Partial support for this work was provided at the University of Minnesota through National Science Foundation grant AST-06-07674.  We appreciate discussions with Bill Cotton (NRAO) regarding NVSS instrumental issues, and Eric Greisen (NRAO) for work on the all sky map combination software.  Bryan Gaensler (Sydney) provided very useful input on an earlier draft. M. Wolleben confirmed and performed a rough fit to our possible new galactic loop, and J. Stil provided useful perspectives on the noise characteristics.

The Very Large Array is a facility of the National Science Foundation, operated by NRAO under contract with AUI, Inc. The Dominion Radio Astrophysical Observatory is operated as a national Facility by the National Research Council Canada.  We also acknowledge the use of NASA's SkyView facility  (http://skyview.gsfc.nasa.gov) located at NASA Goddard Space Flight Center.
\end{acknowledgements}

%begin{deluxetable}{cccccccc}
%tabletypesize{\scriptsize}
%tablecaption{Properties of Giant Radio Galaxies \label{table1}}
%tablewidth{0pt}
%tablehead{
%colhead{Source} & \colhead{RA} & \colhead{Dec} & \colhead{\p800 flux} & \colhead{Ang. Size}  & \colhead{z} & %colhead{\p800 Extent}  \\
%colhead {} &  \colhead{(J2000)} & \colhead{(J2000)} & \colhead{(mJy/beam)}  & \colhead{(arcmin)} & \colhead{} & \colhead{Mpc}
%
%startdata

%GC315 & 00 12 34 & 56 78 90 & 1.97E-002 & 7.90E-002 & 8.36427E+35 & 3.34571E+36 & 6.2  \\

%enddata
%end{deluxetable}

\end{document}